\documentclass[twocolumn]{aastex62}

\begin{document}

\title{Yes, multi-periodic dwarfs in Upper Scorpius are binaries}

\author{Andrei Tokovinin}
\affiliation{Cerro Tololo Inter-American Observatory, Casilla 603, La Serena, Chile}
\author{Cesar Brice\~no}
\affiliation{Cerro Tololo Inter-American Observatory, Casilla 603, La Serena, Chile}

\email{atokovinin@ctio.noao.edu}
\correspondingauthor{Andrei Tokovinin}

\begin{abstract}
We found that multi-periodic low-mass stars discovered by {\it Kepler}
K2 in the  Upper Scorpius association are close  binaries with typical
separations of the  order of 10 au and large  mass ratios. These stars
were  surveyed by speckle  interferometry at  the SOAR  telescope with
spatial  resolution of  0\farcs04.  Out  of 129  observed  targets, we
resolved  70  pairs (including 16  previously known  ones and
three new triple systems).   The distribution of projected separations
of binaries with  primary stars less massive than  the Sun corresponds
to a log-normal  with median of 11.6 au  and logarithmic dispersion of
0.60 dex,  similar to M dwarfs  in the field.  Future  orbits of newly
discovered binaries  will provide  accurate measurements of  masses to
calibrate pre-main sequence evolutionary  tracks; a tentative orbit of
one previously known binary is determined here.
\end{abstract}


\section{Introduction}
\label{sec:intro}

As stars condense from gas, the excess of angular momentum is removed
by a combination of mechanisms. Rotation and multiplicity of young
stars reflect the result of this complex, still poorly understood
process where disks play a major role. This is the context of the
present work.

The  {\it Kepler}  K2  campaigns furnished  massive   amounts  of
high-quality  photometry   of  several  young   stellar  clusters  and
associations, bringing  statistical studies  of stellar rotation  to a
new level.  \citet{Rebull2018} (hereafter RSC18) found that about 20\%
of low-mass  stars in  the Pleiades and  Praesepe clusters and  in the
Upper  Scorpius  (USco)  association  have  two  or  more  photometric
periods.   They  interpreted this  as  different  rotation periods  of
comparable-brightness components in unresolved binaries.  The location
of  multi-periodic  stars   on  the  color-magnitude  diagrams  (CMDs)
supports  this  interpretation.  However,  Rebull  et  al.  note  that
multiple periods can result  from other phenomena such as differential
rotation;  they   believe  that  low-mass   multi-periodic  stars  are
predominantly binaries, while some multi-periodic stars of higher mass
are not  binary.  On the other  hand, a binary with  a large magnitude
difference  or an  inactive component  may have  only  one photometric
period.   So,  not  all   binaries  are  multi-periodic  and  not  all
multi-periodic stars are binary.
  
Identification  of most  low-mass multi-periodic  stars  with binaries
still  lacks  a direct  proof.   Moreover,  the  separations of  those
hypothetical binaries  remain unknown.  Binary  separation impacts the
size  and survival  of  the circumstellar  disks  which influence  the
rotation.    \citet{Stauffer2018}  determined   that   young  low-mass
multi-periodic stars rotate, on average, faster than single-stars, and
related this  finding to the  different disk properties in  single and
multiple stars.

We  observed 129  multi-periodic stars  in USco   from  Table~1 of
  RSC18 by speckle interferometry  and spatially resolved 70 of them.
This   supports  the   proposed  interpretation   of     low-mass
multi-periodic  stars  as  being  mostly  binaries  and  provides  the
distribution  of  their  separations.   Most  resolved  binaries  have
components of  similar brightness, but  we cannot tell to  what extent
this reflects the  distribution of the mass ratios  because our sample
favors near-equal binaries.

The observations are presented in Section~\ref{sec:obs}, their results
are   given   in    Section~\ref{sec:res}   and   are   discussed   in
Section~\ref{sec:disc}.

\section{Observations}
\label{sec:obs}

\subsection{Instrument and data}

We  used the  high-resolution camera  (HRCam)  on the  4.1 m  Southern
Astrophysical Research  Telescope (SOAR) located at  Cerro Pach\'on in
Chile. The  detector in HRCam has  been recently replaced  by a better
camera, gaining at least one magnitude in sensitivity. The instrument,
observing procedure, and data  reduction are covered in \citep{HRCAM};
see recent results and references in \citep{SAM18}.

This project did  not have regular time allocation  at SOAR.  However,
we observed these targets on 2018  April 2 for two hours while working
with HRCam on  another project.  Our goal was  to evaluate quickly the
feasibility of this study. Surprisingly, we could cover 52 targets and
resolved 2/3  of them.  This  provided a strong stimulus  to continue.
Another two  hours were devoted to  this work on 2018  May 26, enlarging
the observed sample.  The third and last observing run on 2018 June 25
used two hours of engineering time.

The  alt-azimuth mount  of  SOAR prevents  observations  close to  the
zenith.   So,  the  USco  region (declination  $\sim  -25^\circ$)  was
observed before  the meridian,  on the Eastern  side. On April  2, the
seeing was  excellent, the vibrations  inherent to the  SOAR telescope
were absent,  and we  obtained good data  on the 52  brightest targets
with $V <  15.5$ mag. As these observations  were exploratory, we took
for  each  target   only  two  data  cubes:  one   with  the  standard
200$\times$200 pixel (3\farcs15 size)  format and 50 ms exposure time,
another  with  the  extended  400$\times$400  pixel  size,  2$\times$2
detector binning, and a longer  exposure of 100\,ms.  This second data
cube  would detect fainter  companions at  separations up  to 3\arcsec
~(which are  rare in this  sample). All data  were taken with  the $I$
filter  (824/170\,nm bandwidth).   

On 2018 May 26, the seeing was  not so good. We pursued the project by
observing 32 targets down to  $V=16$ mag. Fainter magnitudes and worse
seeing degraded the  sensitivity, and we had to  use the long exposure
time of  50\,ms and the 2$\times$2  binning. For each  target,  two
200$\times$200 (narrow-field) data cubes  were taken, so we could miss
companions  wider than  1\farcs5.  The  resolution was  somewhat worse
than  on the  first night,  so we  could have  also missed  some close
pairs.

The last observing run on June  25 had an average seeing. We took data
in the standard mode (without  binning) for  targets brighter than
  $I=12$ mag and  with 2$\times$2 binning for fainter  ones.  In this
run, we  extended the  program to multi-periodic stars with $I<10$
  mag  from RSC18,  thus hoping  to  enlarge the  number of  pre-main
sequence (PMS)  binaries suitable  for future orbit  determination and
measurement of masses.

The data  cubes were  processed in the  standard way  \citep{HRCAM} by
calculating   the   power   spectra   and   the   associated   speckle
auto-correlation functions  (ACFs). The shift-and-add  images centered
on the brightest  pixel in each frame are  produced as well.  However,
they  are useful  only  for  determination of  the  true quadrant  for
brighter targets, when the companion  is detectable in such images and
the magnitude difference exceeds 0.5 mag.  Parameters of each resolved
binary  (position  angle, separation,  and  magnitude difference)  are
determined by fitting a model  to the power spectrum.  The pixel scale
(0\farcs01575) and  detector orientation are accurately  measured on a
set of calibration binaries for  each observing run.

\begin{figure}
\epsscale{1.1}
\plotone{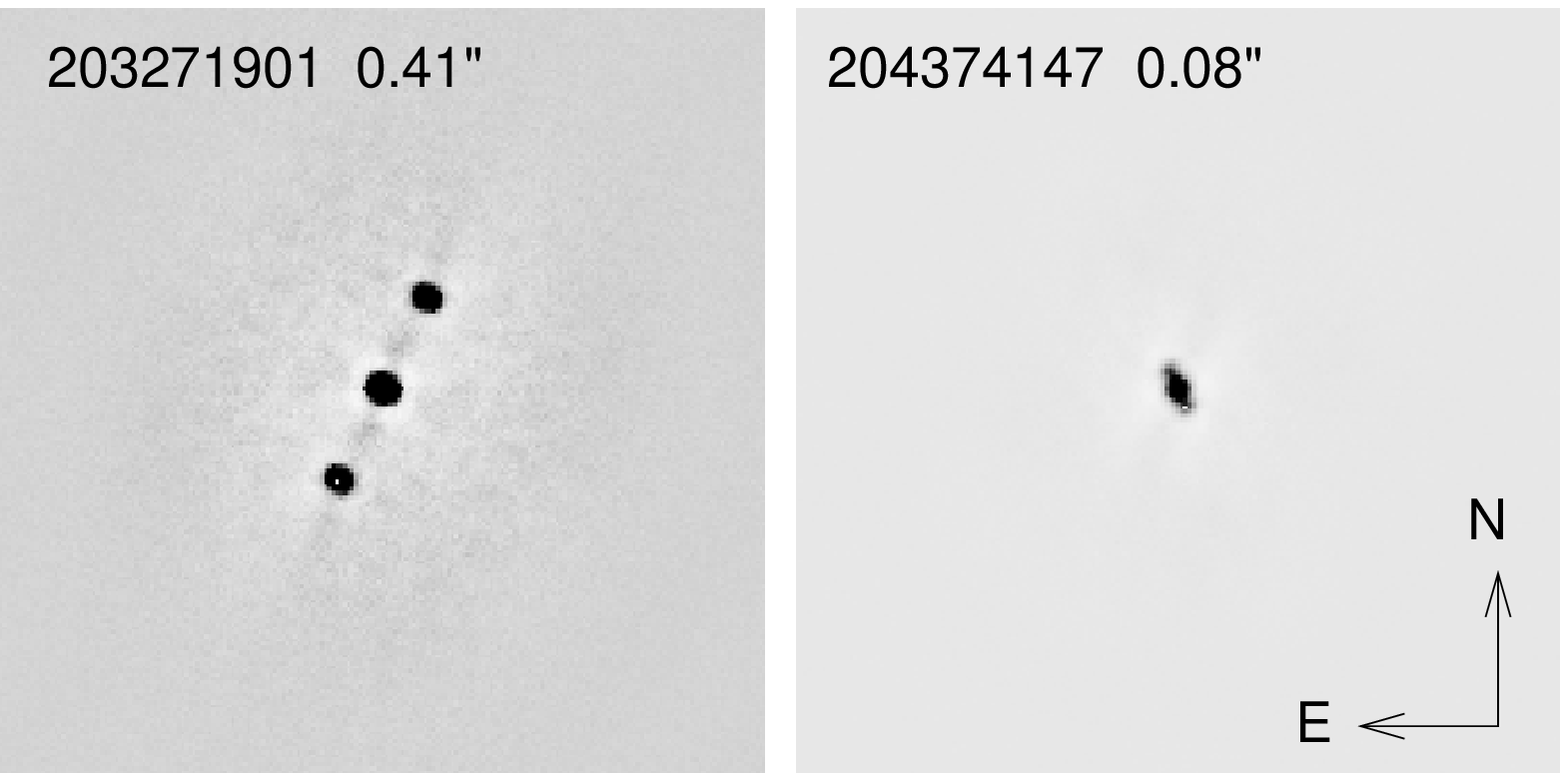}
\plotone{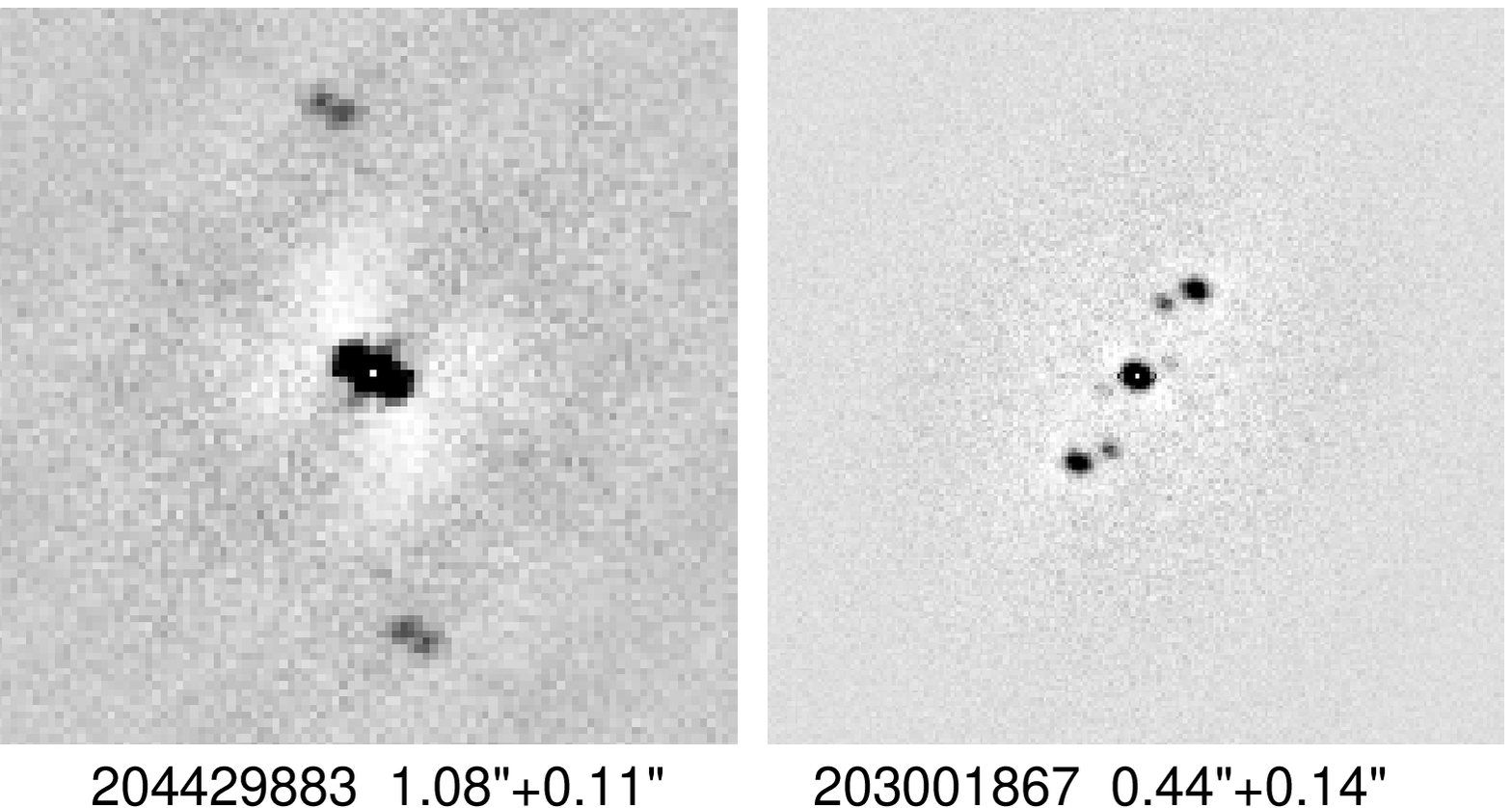}
\caption{Speckle  auto-correlation  functions  of two  newly  resolved
  binaries   (top)  and   two  new   triples  (bottom),   in  negative
  rendering.  The  frame size  is  3\farcs15,  North  is up  and  East
  left. The EPIC numbers and separations are indicated.
\label{fig:ACF}
}
\end{figure}

Figure~\ref{fig:ACF} gives examples of the ACFs for two resolved
binaries and two newly discovered triple systems. 

\subsection{Observed sample}

\begin{figure}
\epsscale{1.1}
\plotone{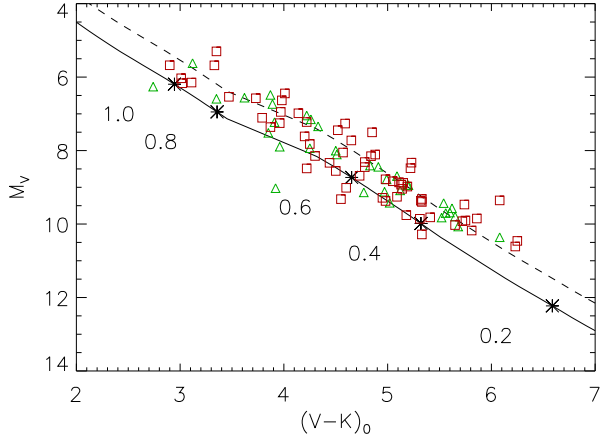}
\caption{Color-magnitude  diagram  for  the sample  of  multi-periodic
  dwarfs  in  USco observed  here    (absolute  $V$ magnitude  vs.
    de-reddened  color  $(V-K)_0$).   Red  squares mark  the  resolved
    binaries, green triangles are  unresolved stars. The full line is
  the  8  Myr  isochrone  for solar  metallicity  from  \citet{PARSEC}
  (asterisks and  numbers indicate masses), the dashed  line shifts it
  up by 0.75 mag.
\label{fig:CMD}
}
\end{figure}

 Table~1 of RSC18 contains 239 multi-periodic stars in USco.  The
sensitivity of HRcam allowed us  to observe only the brightest part of
 this sample  with  $I  <  13$  mag  and  $(V-K)_0  <  6$  mag,
corresponding  to  spectral  types  earlier than  M5V,  approximately.
Figure~\ref{fig:CMD}  shows  the $(M_V,  V-K)$  CMD  for the  observed
targets.  We used the {\it Gaia} DR2 \citep{Gaia} parallaxes (replaced
by the  average parallax  of 7.04\,mas if  not available)  and applied
individual de-reddening  corrections  from RSC18.   The full line
is a 8  Myr isochrone for solar metallicity  from \citet{PARSEC}, also
used by Rebull  et al.  As all these stars are  deemed to be binaries,
they should be displaced from the isochrone upwards by as much as 0.75
mag (dashed line).  On average, the observed stars are located 0.5 mag
above the isochrone, with an rms scatter of 0.7 mag.

The  masses of  main components  were estimated  from  the de-reddened
absolute $V$  magnitudes by  assuming the average  shift of  +0.75 mag
above the 8 Myr isochrone.  Considering the slope of the isochrone and
the scatter, the  errors of the estimated masses  are $\sim$0.1 ${\cal
  M}_\odot$.  Although these masses depend on the model isochrones and
assumed age  and hence could be  biased, they are  useful for relative
ranking of our objects in mass.    The masses range from 0.25 to 3
  ${\cal M}_\odot$;  the median mass  is 0.63 ${\cal  M}_\odot$.  The
colors $(V-K)_0 > 3$ mag  correspond to spectral types later than K5V,
but these  PMS stars  will have somewhat  earlier spectral  types when
they arrive on the main sequence.

\startlongtable
\begin{deluxetable*}{c cc ccc ccc l }
\tabletypesize{\scriptsize}     
\tablecaption{The observed sample
\label{tab:sample} }  
\tablewidth{0pt}                                   
\tablehead{      
\colhead{EPIC} & 
\colhead{$\alpha$} & 
\colhead{$\delta$} & 
\colhead{$I_C$} & 
\colhead{$V-K$} & 
\colhead{$\overline{\omega}$} & 
\colhead{$\rho$} & 
\colhead{$\Delta I$ } & 
\colhead{Mass} & 
\colhead{Note\tablenotemark{a}} \\
& \multicolumn{2}{c}{ J2000} &
\colhead{(mag)} & 
\colhead{(mag)} & 
\colhead{(mas)} & 
\colhead{(\arcsec)} & 
\colhead{(mag)} & 
\colhead{(${\cal M}_\odot$)} & 
}
\startdata
204562176 &  15 55  2.13  & $-$21 49 43.4 & 10.62 &  3.82 &   6.17 & \ldots & \ldots &  0.95 &    \\
204175508 &  15 55 17.58  & $-$23 22  3.7 &  7.92 &  1.04 &   8.24 & \ldots & \ldots &  1.63 &  W?,4  \\
203553934 &  15 55 29.81  & $-$25 44 49.9 & 11.47 &  4.79 & \ldots &   0.37 &    0.2 &  0.59 &  D?  \\
203710077 &  15 55 48.81  & $-$25 12 24.0 &  9.52 &  2.21 &   6.95 & \ldots & \ldots &  1.45 &    \\
204918279 &  15 56 25.09  & $-$20 16 16.2 & 12.60 &  6.67 & \ldots &   0.18 &    0.2 &  0.29 &    \\
204104740 &  15 56 27.62  & $-$23 38 52.1 & 11.55 &  5.19 &   7.14 &   0.16 &    0.1 &  0.56 &    \\
204832936 &  15 56 42.43  & $-$20 39 34.2 & 12.86 &  5.74 &   7.00 &   1.41 &    3.2 &  0.38 &  D,G  \\
204248842 &  15 57  3.67  & $-$23 04 48.6 & 12.83 &  5.93 &   6.46 & \ldots & \ldots &  0.32 &    \\
203696502 &  15 57 15.56  & $-$25 15 13.3 &  7.99 &  0.87 &   6.20 &   0.20 &    2.0 &  1.73 & W   \\
203628765 &  15 57 16.74  & $-$25 29 19.3 & 10.84 &  3.81 & \ldots &   0.58 &    0.1 &  0.82 & W,D?   \\
204179058 &  15 57 34.31  & $-$23 21 12.3 & 11.27 &  4.63 &   6.93 &   0.05 &    0.6 &  0.74 & W,G+,4   \\
204780592 &  15 58 30.39  & $-$20 53 35.9 &  6.85 &  0.73 &   7.04 & \ldots & \ldots &  2.06 &    \\
204066898 &  15 58 36.21  & $-$23 48  2.1 & 12.70 &  5.47 &   8.17 & \ldots & \ldots &  0.37 &    \\
204281210 &  15 58 36.92  & $-$22 57 15.3 &  8.70 &  3.04 &   6.02 & \ldots & \ldots &  1.52 & D   \\
203301463 &  15 59 11.89  & $-$26 35  0.8 &  7.58 &  1.16 &   7.18 &   0.66 &    1.8 &  1.77 &    \\
204477741 &  15 59 18.38  & $-$22 10 43.2 & 12.74 &  6.19 &   6.77 &   0.05 &    0.0 &  0.31 &    \\
203462615 &  15 59 38.05  & $-$26 03 23.6 & 12.93 &  6.24 &   7.72 & \ldots & \ldots &  0.25 &    \\
203499724 &  15 59 49.88  & $-$25 55 58.8 & 12.01 &  5.04 &   6.29 & \ldots & \ldots &  0.59 &    \\
204436170 &  15 59 59.95  & $-$22 20 36.7 & 10.90 &  4.62 &   7.15 & \ldots & \ldots &  0.72 &  W  \\
204878974 &  16 00 31.35  & $-$20 27  5.0 & 11.16 &  4.79 &  11.04 &   0.14 &    0.4 &  0.56 &  W  \\
204350593 &  16 00 41.34  & $-$22 40 41.8 & 12.16 &  5.50 &   7.40 &   0.11 &    0.1 &  0.42 &    \\
203363279 &  16 00 42.65  & $-$26 23 29.6 &  8.60 &  1.60 &   6.39 & \ldots & \ldots &  1.61 &  3  \\
204406748 &  16 01  5.18  & $-$22 27 31.1 & 11.14 &  4.99 & \ldots &   0.28 &    0.6 &  0.69 &  W  \\
204428324 &  16 01 10.36  & $-$22 22 27.9 & 12.39 &  5.98 &   6.87 & \ldots & \ldots &  0.36 &    \\
203690414 &  16 01 13.97  & $-$25 16 28.4 & 12.95 &  5.54 & \ldots &   0.18 &    0.4 &  0.32 &  D  \\
204794876 &  16 01 47.43  & $-$20 49 45.7 & 10.87 &  4.58 &   6.89 &   0.06 &    0.6 &  0.73 &  W  \\
204246759 &  16 03 33.79  & $-$23 05 19.0 & 10.88 &  3.44 &   6.30 & \ldots & \ldots &  0.93 &    \\
204862109 &  16 03 54.96  & $-$20 31 38.5 & 10.88 &  4.58 &   6.60 &   0.09 &    0.2 &  0.73 &  W  \\
203895983 &  16 04 18.93  & $-$24 30 39.3 & 11.23 &  4.96 & \ldots &   0.29 &    0.1 &  0.74 &  D  \\
204637622 &  16 04 20.98  & $-$21 30 41.6 & 11.99 &  5.68 &   6.79 &   0.06 &    0.0 &  0.50 &  D,d?  \\
204603210 &  16 04 35.87  & $-$21 39 22.3 & 12.42 &  5.68 &   7.29 &   0.09 &    0.6 &  0.39 &  G+  \\
204856827 &  16 06 59.37  & $-$20 33  4.7 & 12.13 &  5.80 &   7.32 &   0.07 &    0.0 &  0.44 &    \\
204844509 &  16 07  3.56  & $-$20 36 26.4 & 10.20 &  4.13 & \ldots &   0.21 &    0.4 &  0.78 &  W   \\
204506777 &  16 07 17.78  & $-$22 03 36.5 &  7.99 &  1.70 &   6.87 & \ldots & \ldots &  1.67 &    \\
204242152 &  16 07 23.75  & $-$23 06 24.1 & 12.33 &  5.32 &   6.33 &   0.07 &    0.0 &  0.45 &    \\
204757338 &  16 07 27.46  & $-$20 59 44.4 & 12.94 &  6.48 & \ldots &   0.57 &    0.0 &  0.30 &  D  \\
204845955 &  16 07 44.48  & $-$20 36  3.2 & 11.66 &  5.74 &   6.77 &   0.06 &    0.4 &  0.53 &  W,3  \\
204229583 &  16 07 57.76  & $-$23 09 23.6 & 11.54 &  5.23 &   7.29 &   2.40 &    1.0 &  0.55 &  G,3  \\
203851147 &  16 07 58.76  & $-$24 41 31.9 & 12.37 &  5.57 & \ldots &   0.64 &    0.6 &  0.38 &    \\
205087483 &  16 08  1.55  & $-$19 27 58.2 & 12.23 &  5.59 & \ldots &   0.84 &    0.0 &  0.46 &  W,G  \\
203271901 &  16 08  4.10  & $-$26 40 44.9 & 11.32 &  4.59 & \ldots &   0.41 &    0.4 &  0.71 &    \\
203788687 &  16 08  5.22  & $-$24 55 33.2 &  8.42 &  1.79 &   6.75 &   0.94 &    2.8 &  1.63 &  W  \\
205080616 &  16 08 23.24  & $-$19 30  0.9 & 11.79 &  4.76 &   7.25 & \ldots & \ldots &  0.60 &  D  \\
205141287 &  16 08 34.36  & $-$19 11 56.2 &  9.81 &  3.95 & \ldots &   0.31 &    1.1 &  1.03 &  3  \\
204810792 &  16 08 35.14  & $-$20 45 29.6 &  7.60 &  1.67 &   7.89 & \ldots & \ldots &  1.68 &    \\
204099739 &  16 08 39.08  & $-$23 40  5.6 & 11.63 &  5.33 &   6.92 & \ldots & \ldots &  0.56 &    \\
205177770 &  16 08 43.07  & $-$19 00 52.1 & 12.91 &  6.05 & \ldots &   0.15 &    0.6 &  0.30 &    \\
204852312 &  16 08 54.07  & $-$20 34 18.4 & 12.69 &  6.33 &   7.06 & \ldots & \ldots &  0.34 &    \\
202724025 &  16 08 56.94  & $-$28 35 57.7 & 12.19 &  5.67 &   6.85 & \ldots & \ldots &  0.70 &    \\
204429883 &  16 09 20.62  & $-$22 22  5.9 & 12.20 &  6.40 & \ldots &   0.12 &    0.0 &  0.38 &  W,T,G,4  \\
205203376 &  16 09 29.19  & $-$18 52 53.7 & 10.56 &  4.36 &  10.63 &   0.22 &    1.6 &  0.69 &    \\
204608292 &  16 09 35.74  & $-$21 38  5.9 & 12.38 &  5.29 &  11.00 &   0.12 &    1.3 &  0.36 &    \\
204059992 &  16 10  1.83  & $-$23 49 43.5 & 12.72 &  5.79 &   7.10 & \ldots & \ldots &  0.58 &    \\
203036995 &  16 10  3.09  & $-$27 28 39.8 & 12.39 &  5.62 & \ldots &   0.12 &    0.7 &  0.38 &    \\
204060981 &  16 10 56.18  & $-$23 49 29.2 & 12.06 &  5.80 &   6.84 & \ldots & \ldots &  0.47 &    \\
203716047 &  16 10 57.91  & $-$25 11 10.3 & 12.43 &  5.40 & \ldots &   0.46 &    0.6 &  0.39 &    \\
203754361 &  16 11  7.43  & $-$25 03  1.8 & 12.12 &  5.43 &   6.18 & \ldots & \ldots &  0.49 &    \\
204217530 &  16 11 15.95  & $-$23 12 14.6 &  8.43 &  1.24 &   6.46 & \ldots & \ldots &  1.60 & D,4   \\
205168266 &  16 12  5.99  & $-$19 03 44.3 &  7.44 &  1.19 &   7.19 & \ldots & \ldots &  2.13 &    \\
204857023 &  16 12  8.24  & $-$20 33  1.6 & 12.46 &  5.94 &   7.05 &   1.80 &    0.6 &  0.54 &    \\
203760219 &  16 13  2.34  & $-$25 01 46.0 & 11.69 &  4.57 &   7.12 &   0.07 &    0.2 &  0.69 &    \\
205373716 &  16 13 20.80  & $-$17 57 52.1 & 12.47 &  5.83 &   6.99 & \ldots & \ldots &  0.35 &    \\
205373893 &  16 13 21.23  & $-$17 57 49.1 & 12.61 &  5.67 &   7.37 & \ldots & \ldots &  0.40 &    \\
203048597 &  16 13 24.56  & $-$27 26 13.4 & 12.31 &  5.52 & \ldots &   0.74 &    0.0 &  0.39 &  G  \\
204156820 &  16 13 36.44  & $-$23 26 27.0 & 11.74 &  5.14 & \ldots &   0.55 &    0.0 &  0.59 &    \\
205225696 &  16 13 43.50  & $-$18 45 52.8 & 13.26 &  6.64 & \ldots &   0.21 &    0.0 &  0.34 &    \\
204459941 &  16 13 43.66  & $-$22 14 59.4 & 11.20 &  4.10 &   6.27 &   0.06 &    0.4 &  0.73 &    \\
203917770 &  16 13 45.49  & $-$24 25 19.5 &  6.14 &  0.35 &   5.98 &   1.74 &    3.0 &  2.97 & W,G,3   \\
202947197 &  16 13 47.81  & $-$27 47 33.9 &  8.69 &  1.28 &   6.57 &   0.36 &    4.5 &  1.54 & T,D?,4   \\
203777800 &  16 13 56.62  & $-$24 57 56.9 & 12.65 &  5.42 & \ldots &   0.20 &    0.1 &  0.39 &    \\
203799762 &  16 14  3.79  & $-$24 53  9.1 & 12.29 &  4.60 &   6.48 & \ldots & \ldots &  0.60 &    \\
204449800 &  16 14 10.10  & $-$22 17 23.7 & 12.58 &  6.91 &   6.46 &   0.84 &    0.0 &  0.31 &  G  \\
205047378 &  16 14 13.81  & $-$19 39 36.2 &  8.56 &  2.36 &   6.28 & \ldots & \ldots &  1.55 &    \\
204488355 &  16 14 20.50  & $-$22 08  9.7 &  7.28 &  0.84 &   7.25 & \ldots & \ldots &  1.78 &  4  \\
205188906 &  16 14 28.92  & $-$18 57 22.5 & 11.92 &  5.23 &   7.08 & \ldots & \ldots &  0.51 &  D  \\
204235325 &  16 14 52.68  & $-$23 08  2.7 & 12.11 &  5.44 &   7.13 &   0.70 &    1.2 &  0.44 &    \\
202533810 &  16 15  0.60  & $-$29 19 34.8 & 11.80 &  4.91 & \ldots &   0.17 &    0.0 &  0.52 &    \\
204204606 &  16 15 54.86  & $-$23 15 14.9 & 12.86 &  5.52 & \ldots &   0.27 &    0.0 &  0.33 &    \\
203433962 &  16 16 17.20  & $-$26 09 10.2 & 10.96 &  4.51 &   6.50 & \ldots & \ldots &  0.72 &    \\
204082531 &  16 16 20.10  & $-$23 44 14.4 & 12.78 &  5.81 &   7.91 &   0.06 &    0.0 &  0.27 &    \\
203750949 &  16 16 22.94  & $-$25 03 46.7 &  7.84 &  1.17 &   6.48 & \ldots & \ldots &  1.77 &    \\
203771564 &  16 16 25.17  & $-$24 59 19.4 &  7.47 &  0.91 &   6.37 & \ldots & \ldots &  2.31 & W?   \\
203703588 &  16 16 44.25  & $-$25 13 45.3 &  8.32 &  1.15 &   6.85 & \ldots & \ldots &  1.65 & 4   \\
203855509 &  16 16 47.93  & $-$24 40 28.4 & 12.93 &  6.18 &   6.35 &   0.07 &    0.4 &  0.35 & 3   \\
203809634 &  16 17 26.14  & $-$24 50 59.4 & 12.49 &  6.02 &   6.58 & \ldots & \ldots &  0.43 &    \\
203809317 &  16 17 29.94  & $-$24 51  3.2 & 12.36 &  5.71 & \ldots &   0.41 &    0.5 &  0.52 &    \\
205267399 &  16 18 41.87  & $-$18 32 40.1 & 12.65 &  5.70 &   8.67 & \ldots & \ldots &  0.29 &    \\
204569229 &  16 19 45.38  & $-$21 47 57.7 & 11.27 &  4.38 &   7.54 &   0.05 &    0.4 &  0.66 &    \\
204655550 &  16 20 27.24  & $-$21 26  6.8 & 11.41 &  4.45 &   7.78 &   0.08 &    0.9 &  0.63 &    \\
204666965 &  16 20 36.40  & $-$21 23 12.3 & 12.62 &  5.86 &   7.39 & \ldots & \ldots &  0.31 &    \\
203891936 &  16 20 44.68  & $-$24 31 38.3 & 10.54 &  3.89 &   6.09 & \ldots & \ldots &  0.76 &    \\
204372172 &  16 20 50.23  & $-$22 35 38.7 &  8.44 &  1.24 &   6.95 & \ldots & \ldots &  1.65 & D   \\
204494885 &  16 21 21.14  & $-$22 06 32.2 &  8.02 &  1.22 &   6.29 & \ldots & \ldots &  1.75 &    \\
204471912 &  16 21 41.26  & $-$22 12  5.7 & 12.55 &  5.91 &   7.53 & \ldots & \ldots &  0.33 &    \\
203495721 &  16 21 47.92  & $-$25 56 47.1 &  8.75 &  1.34 &   7.62 & \ldots & \ldots &  1.47 &    \\
204603511 &  16 22 10.05  & $-$21 39 17.3 &  8.15 &  1.00 &   7.19 & \ldots & \ldots &  1.65 &  D,3  \\
202800875 &  16 22 13.14  & $-$28 18 58.5 & 11.21 &  4.14 &   6.30 & \ldots & \ldots &  0.72 &    \\
203517602 &  16 22 22.30  & $-$25 52 20.2 & 12.07 &  6.00 &   6.99 & \ldots & \ldots &  0.51 &    \\
204374147 &  16 22 34.21  & $-$22 35 12.3 & 11.05 &  5.46 &   6.97 &   0.08 &    0.9 &  0.71 &  3  \\
204591415 &  16 22 44.06  & $-$21 42 22.4 & 12.44 &  5.78 &   6.63 & \ldots & \ldots &  0.43 &    \\
202908892 &  16 22 51.57  & $-$27 55 37.4 & 12.63 &  5.22 &   6.83 & \ldots & \ldots &  0.39 &    \\
203001867 &  16 23 59.01  & $-$27 36  3.7 & 11.91 &  5.34 & \ldots &   0.44 &    0.3 &  0.53 &  T  \\
203649927 &  16 24  2.89  & $-$25 24 53.9 &  8.20 &  2.70 &   5.82 & \ldots & \ldots &  2.62 &  D,G+  \\
203783190 &  16 24  6.33  & $-$24 56 46.9 & 10.58 &  4.71 & \ldots &   0.24 &    0.2 &  0.94 &    \\
203761347 &  16 24 21.31  & $-$25 01 31.4 &  7.42 &  2.21 &   7.15 &   2.71 &    4.6 &  2.70 &  W,G,4  \\
204355491 &  16 24 51.36  & $-$22 39 32.5 &  8.58 &  2.73 &   7.22 &   0.04 &    0.5 &  1.57 &  W  \\
204462113 &  16 24 55.33  & $-$22 14 28.6 & 12.69 &  5.30 &   7.91 &   0.08 &    0.2 &  0.40 &    \\
203888630 &  16 25  7.32  & $-$24 32 27.2 & 11.95 &  6.51 & \ldots &   0.29 &    0.0 &  0.84 & 4   \\
203911473 &  16 25 19.24  & $-$24 26 52.7 &  9.05 &  2.26 &   7.45 &   0.14 &    0.0 &  1.57 & W   \\
203071614 &  16 25 55.40  & $-$27 21 24.4 & 12.55 &  5.72 &   8.06 &   0.11 &    0.4 &  0.29 &    \\
203476597 &  16 25 57.91  & $-$26 00 37.3 & 10.79 &  4.58 &   6.32 & \ldots & \ldots &  0.77 &    \\
203086979 &  16 26  2.15  & $-$27 18 14.1 & 11.02 &  3.79 &   7.47 & \ldots & \ldots &  0.73 &    \\
203881373 &  16 26  9.31  & $-$24 34 12.1 &  9.01 &  5.45 &   7.06 & \ldots & \ldots &  2.88 &  D  \\
203937317 &  16 26 17.07  & $-$24 20 22.0 & 10.72 &  6.15 &   7.45 & \ldots & \ldots &  0.93 &  D  \\
204611221 &  16 26 19.61  & $-$21 37 21.0 & 11.84 &  5.13 &   7.53 & \ldots & \ldots &  0.55 &    \\
204520585 &  16 26 41.19  & $-$22 00  9.7 & 11.75 &  5.06 &   5.90 &   0.11 &    0.9 &  0.70 &    \\
203856244 &  16 26 41.22  & $-$24 40 17.7 & 12.11 &  7.95 &   5.96 &   0.14 &    0.5 &  0.44 & W,D   \\
203442191 &  16 27  6.68  & $-$26 07 31.1 &  9.46 &  2.11 &   6.14 & \ldots & \ldots &  1.57 &    \\
203850605 &  16 27 19.49  & $-$24 41 40.3 & 10.76 &  4.87 &   8.90 &   0.15 &    0.0 &  0.60 & W,D   \\
203115615 &  16 28 45.14  & $-$27 12 19.5 & 12.36 &  5.59 &   4.55 &   0.13 &    1.0 &  0.65 &    \\
202615424 &  16 29 11.39  & $-$29 00 31.7 & 12.34 &  5.00 & \ldots &   0.13 &    0.0 &  0.52 &    \\
203873374 &  16 29 35.09  & $-$24 36 10.7 & 12.50 &  8.84 & \ldots &   0.19 &    0.1 &  0.48 &  D  \\
204559470 &  16 33 38.82  & $-$21 50 26.3 & 11.38 &  5.47 &   6.78 & \ldots & \ldots &  0.71 &    \\
203181641 &  16 34  5.85  & $-$26 58 44.2 & 10.67 &  3.60 &   7.32 &   0.09 &    1.1 &  0.87 &    \\
202679988 &  16 35 11.88  & $-$28 45 52.0 & 10.93 &  3.72 &   6.67 &   0.06 &    0.9 &  0.84 &    \\
205059490 &  16 36 39.22  & $-$19 36  7.5 &  8.82 &  1.32 &   6.24 & \ldots & \ldots &  1.61 &  4  \\
203135082 &  16 36 52.88  & $-$27 08 18.6 &  9.23 &  2.04 &   6.21 & \ldots & \ldots &  1.47 &    \\
205075874 &  16 38 28.57  & $-$19 31 26.0 &  9.31 &  1.83 &   5.03 & \ldots & \ldots &  1.64 &    \\
205002311 &  16 47 47.33  & $-$19 52 31.9 &  8.47 &  1.26 &   8.46 & \ldots & \ldots &  1.48 &  D  
\enddata
\tablenotetext{a}{Notes: W -- listed in WDS; D -- disk or possible
  disk in RSC18; G --
  companion measured in Gaia; G+ -- additional Gaia companion; T -- triple; 3.4 -- three or 4 periods.}
\end{deluxetable*}
   
Table~\ref{tab:sample}  lists the  129 observed  stars,  identified by
their EPIC numbers in column  (1) and J2000 coordinates in columns (2)
and (3).  In  columns (4) and (5) we give the  $I_C$ magnitude and the
$(V-K)$ color (without reddening  correction), as determined by RSC18.
Column (6) contains the  {\it Gaia} DR2 \citep{Gaia} parallaxes, where
available.   The   separations   and  magnitude
differences  of resolved  pairs are  listed  in columns  (7) and  (8),
respectively.  Column  (8) gives the  primary mass estimated  from the
isochrone.  The notes in column (9) are explained below.

We matched the  targets to the {\it Gaia}  DR2 within 6\arcsec ~radius
to get  their parallaxes.  These are  missing only for   28 stars.
  All  these 28 stars are  binaries with  separations from  0\farcs11 to
0\farcs84,  where the  resolved nature  of the  source  was apparently
detected by {\it Gaia}  and prevented parallax measurement. 
Closer binaries are unresolved by {\it Gaia}  and treated there as
  single stars, while components of wider binaries  have separate
  entries in DR2, allowing us  to compute their relative positions and
  magnitdue differences.  {\it Gaia}   measured nine pairs wider
than 0\farcs8,  also measured here.  The agreement  between {\it Gaia}
and  HRcam astrometry  of common  pairs is  very good.   Three targets
(EPIC~204179058,  204603210,  203649927)  have additional  {\it  Gaia}
companions with  matching parallaxes and proper motions  (marked G+ in
Table~\ref{tab:sample})  at separations  of 3\farcs76,  5\farcs52, and
2\farcs06,  respectively.  The first  two stars   also  have close
  companions, so these systems  are in fact triple.  These additional
{\it Gaia}  companions are faint  and fall outside the  4\arcsec ~{\it
  Kepler} pixels, explaining why only double periods were detected for
these stars  by RSC18.  The first one, EPIC~204179058, is flagged
by RSC18 as potentially disk-bearing.  EPIC~204248842 has a {\it Gaia}
source  at 1\farcs2 separation  with a  large magnitude  difference of
$\Delta G =  5.1$ mag and unknown parallax; we  have not detected this
companion  with  HRcam and  consider  it  optical  or spurious.   
  Finally,  two  objects (EPIC  204878974  and  20408292)   have
parallaxes  of 11\,mas,   challenging their  membership in  USco;
  however, both are 0\farcs1 binaries unresolved by {\it Gaia}, hence
these parallaxes  are likely biased.   The {\it  Gaia} astrometry
  confirms membership of all our targets in the USco association.

The  sample was  created regardless  of known  binaries listed  in the
Washington Double Star Catalog, WDS \citep{WDS}.  In fact, 16 resolved
targets are previously known binaries (marked by W in the notes).  One
previously  known binariy  from this  sample was  not detected  by us,
being   too  close  (EPIC   204436170,  WDS   J16000$-$2221,  KSA~122,
0\farcs025);  two more (EPIC  204175508 =  J15553$-$2322 =  KOH~25 and
203771564 = J16164$-$2459  = KOU~59) with separations on  the order of
1\arcsec  ~are not  confirmed either  by us  or by  {\it  Gaia}, being
apparently spurious  (marked W?).  Overall,  we resolved 57  new pairs
(counting each  subsystem in triples separately); the  total number of
resolved  pairs in our sample  stands at 70. The other symbols in
the notes  are T -- triple,  D or D?   -- star with disk  according to
RSC18, 3 or 4 -- number of periods detected, if more than two.

\begin{figure}
\epsscale{1.1}
\plotone{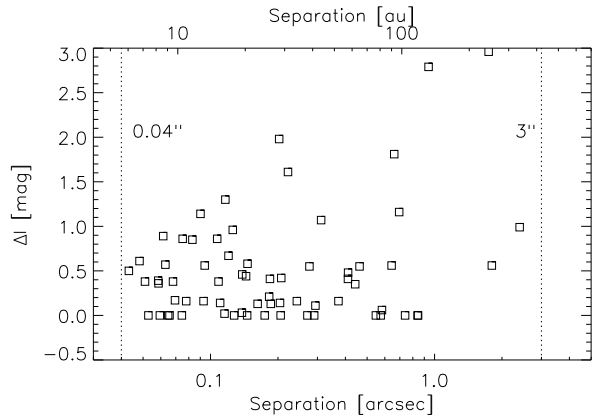}
\caption{Magnitude  difference $\Delta I$  vs. angular  separation for
  resolved binaries.   The upper axis assumes distance  of 140\,pc for
  converting angular separations to au.  Vertical dotted lines delimit
  the surveyed range of separations.
\label{fig:sep-dm}
}
\end{figure}

Figure~\ref{fig:sep-dm}  plots  magnitude  difference $\Delta  I$  vs.
separation  to characterize  the  limits of  our observations.   Small
separations  and moderate  $\Delta I$  dominate, with  a  sharp cutoff
below the  resolution limit of  0\farcs04.  The plot hints  that there
are many even closer, unresolved  binaries in our sample. On the other
hand,  only  4  pairs  out  of 70  have  separations  above  1\farcs5,
exceeding the  maximum detectable separation in  the narrow-field data
cubes.  All  these (and  wider) companions are  also resolved  by {\it
  Gaia},  so our  survey is  not impaired  by the  narrow  field size.
Moreover, the vast majority of pairs are closer than 1\farcs5.

\section{Results}
\label{sec:res}

\subsection{Astrometry} 

Table~\ref{tab:double} gives measurements of resolved pairs. Its first
column  has the  WDS-style code  based on  the J2000  coordinates. For
previously  known pairs, their WDS discoverer codes  are given  in column
(2), while column (3) gives the EPIC numbers. Then follow the date of
observation (4), position angle (5), and the measurement error in
tangential direction $\sigma_\theta$ in mas (6). Columns (7) and (8)
give the separation and its error, respectively. The last column (9)
gives $\Delta I$ to within 0.1 mag, which is the minimum error of
differential speckle photometry. The flags in this column mean noisy
data (:), alternative photometry of wider pairs  (*), and quadrant
identification (q).  

For  nine resolved  pairs, {\it  Gaia} DR2  astrometry  gives relative
positions  and  magnitude differences  in  the  $G$  band (flag  G  in
Table~\ref{tab:sample}).      We     include     those     data     in
Table~\ref{tab:double}, with the epoch of 2015.5 and the flag G in the
last  column.  The  {\it Gaia}  results give  independent test  of our
measurements.  We also  include positions  of three  wide  G+ physical
companions with discovery code 'Gaia'.

\startlongtable

\begin{deluxetable*}{l l l  c  cc cc l  }  
\tabletypesize{\scriptsize}   
\tablecaption{Measurements of resolved pairs       
\label{tab:double} }                                                             
\tablewidth{0pt}       
\tablehead{  
\colhead{WDS} & \colhead{Discoverer} & \colhead{EPIC} & \colhead{Epoch}  & \colhead{$\theta$}     & \colhead{$\sigma_\theta$}  &       
\colhead{$\rho$} & \colhead{$\sigma \rho$} & \colhead{$\Delta I$}          \\
\colhead{(2000)} & \colhead{Designation} &  & +2000 & \colhead{(deg)} & (mas) & ($''$) & (mas) & (mag)                    
}
\startdata     
15555$-$2545 &	\ldots &	203553934 &	18.2524 &	137.4 &	0.3 &	0.3725 &	0.3 &	0.2      \\ 
15564$-$2016 &	\ldots &	204918279 &	18.3999 &	173.7 &	6.1 &	0.1827 &	3.5 &	0.2 :    \\ 
            &	       &	       &	18.4846 &	177.3 &	1.6 &	0.1695 &	2.2 &	0.0 :    \\ 
15565$-$2339 &	\ldots &	204104740 &	18.2524 &	102.2 &	0.2 &	0.1623 &	0.3 &	0.1      \\ 
15567$-$2040 &	\ldots &	204832936 &	18.4846 &	38.9 &	8.0 &	1.4086 &	8.0 &	3.2 *    \\ 
             &         &                  &     15.5000 &       39.3 & \ldots & 1.407  & \ldots     &   3.2 G \\     
15573$-$2515 &	B 301       &	203696502 &	18.4817 &	256.0 &	0.6 &	0.2025 &	0.2 &	2.0 q    \\ 
15573$-$2529 &	KOH 57 AB &	203628765 &	18.2523 &	143.5 &	0.2 &	0.5826 &	0.4 &	0.1      \\ 
            &	       &	          &	18.3999 &	143.7 &	0.6 &	0.5832 &	0.9 &	0.1      \\ 
15576$-$2321 &	KSA 119 AB &	204179058 &	18.2524 &	87.2 &	1.4 &	0.0482 &	2.3 &	0.6      \\ 
             & Gaia AC     &              &     15.5000 &      350.9 & \ldots & 3.760  &  \ldots    &   6.1 G    \\  
15592$-$2635 &	\ldots       &	203301463 &	18.4817 &	191.9 &	0.3 &	0.6608 &	0.2 &	1.8 *    \\ 
            &	       &	          &	18.4817 &	192.0 &	0.8 &	0.6626 &	1.2 &	3.1      \\ 
15593$-$2211 &	\ldots   &	204477741 &	18.4846 &	84.2 &	18.9 &	0.0528 &	18.9 &	0.0 :    \\ 
16001$-$2027 &	KOH 63 &	204878974 &	18.2526 &	134.6 &	1.5 &	0.1441 &	1.8 &	0.4      \\ 
16007$-$2241 &	\ldots &	204350593 &	18.4847 &	56.6 &	1.4 &	0.1104 &	2.7 &	0.1 :    \\ 
16011$-$2228 &	KOH 65 &	204406748 &	18.2526 &	339.7 &	0.1 &	0.2764 &	0.6 &	0.6 q    \\ 
16012$-$2516 &	\ldots &	203690414 &	18.3999 &	50.7 &	9.3 &	0.1842 &	3.6 &	0.4 :    \\ 
16018$-$2050 &	KOH 67 &	204794876 &	18.2526 &	358.8 &	3.6 &	0.0629 &	2.2 &	0.6      \\ 
16039$-$2032 &	KOH 70 AB &	204862109 &	18.2526 &	91.0 &	1.4 &	0.0931 &	3.5 &	0.2      \\ 
            &	       &	          &	18.3999 &	93.9 &	1.7 &	0.0968 &	0.6 &	0.1 :    \\ 
16043$-$2131 &	KOH 71 &	204637622 &	18.2526 &	139.4 &	0.3 &	0.0647 &	3.5 &	0.0 :    \\ 
16043$-$2431 &	\ldots &	203895983 &	18.4846 &	67.2 &	0.4 &	0.2940 &	0.5 &	0.1      \\ 
16046$-$2139 &	AB     &	204603210 &	18.3999 &	131.7 &	1.1 &	0.0942 &	1.6 &	0.6 :    \\ 
             & Gaia AC &                  &     15.5000 &       93.7  & \ldots & 5.527 & \ldots     &   1.6 G      \\
16070$-$2033 &	\ldots &	204856827 &	18.2526 &	120.7 &	0.3 &	0.0656 &	0.3 &	0.0 :    \\ 
16071$-$2036 &	KSA 125 Aa,Ab &	204844509 &	18.2526 &	339.8 &	0.5 &	0.2069 &	0.6 &	0.4 q    \\ 
16074$-$2306 &	\ldots &	204242152 &	18.2526 &	13.1 &	1.7 &	0.0745 &	1.4 &	0.0 :    \\ 
16075$-$2060 &	\ldots &	204757338 &	18.4847 &	112.5 &	1.7 &	0.5749 &	1.1 &	0.0 :    \\ 
16077$-$2036 &	BOY 20 &	204845955 &	18.2526 &	57.8 &	0.3 &	0.0584 &	0.3 &	0.4 :    \\ 
16080$-$1928 &	LAF 114 &	205087483 &	18.4847 &	313.8 &	0.5 &	0.8443 &	0.5 &	0.0      \\ 
             &          &                 &     15.5000 &       314.0 & \ldots  & 0.845 & \ldots    &   0.1 G \\
16080$-$2309 &	\ldots  &	204229583 &	18.2524 &	35.8 &	0.1 &	2.3983 &	0.1 &	1.0 :    \\ 
             &          &                 &     15.5000 &       35.9 & \ldots&  2.388  & \ldots     &   1.0 G   \\
16080$-$2442 &	\ldots  &	203851147 &	18.3999 &	228.7 &	1.1 &	0.6432 &	1.1 &	0.6 :    \\ 
16081$-$2456 &	DON 781 &	203788987 &	18.4818 &	338.7 &	1.1 &	0.9404 &	0.6 &	2.8 *    \\ 
16081$-$2641 &	\ldots &	203271901 &	18.2524 &	154.0 &	0.2 &	0.4107 &	0.5 &	0.4      \\ 
16086$-$1912 &	\ldots &	205141287 &	18.4847 &	143.0 &	0.6 &	0.3117 &	0.4 &	1.1 q    \\ 
16087$-$1901 &	\ldots &	205177770 &	18.4847 &	141.1 &	1.1 &	0.1463 &	8.8 &	0.6 :    \\ 
16093$-$2222 &	WSI 130 Aa,Ab &	204429883 &	18.3999 &	61.2 &	4.9 &	0.1155 &	0.9 &	0.0 :    \\ 
             &                &           &     15.3354 &       51.0  & \ldots& 0.126  &  \ldots    &   0.2     \\
16093$-$2222 &	WSI 130 Aa,B &	204429883 &	18.3999 &	187.3 &	4.5 &	1.0843 &	6.8 &	1.8 :    \\ 
             &                &           &     15.3354 &       190.2 & \ldots &   1.1458 & \ldots  &   2.2 q    \\  
             &                &           &     15.5000 &       189.5 & \ldots &   1.121  &\ldots   &   2.1 G \\
16095$-$1853 &	\ldots &	205203376 &	18.2526 &	166.2 &	2.5 &	0.2217 &	2.2 &	1.6 q    \\ 
16096$-$2138 &	\ldots &	204608292 &	18.2526 &	131.0 &	3.7 &	0.1165 &	1.0 &	1.3 q    \\ 
16101$-$2729 &	\ldots &	203036995 &	18.2524 &	156.6 &	1.6 &	0.1203 &	3.4 &	0.7      \\ 
16110$-$2511 &	\ldots &	203716047 &	18.2524 &	71.5 &	1.4 &	0.4626 &	0.4 &	0.5 :    \\ 
            &	       &	          &	18.4846 &	71.5 &	1.0 &	0.4624 &	1.1 &	0.2 :    \\ 
16121$-$2033 &	\ldots &	204857023 &	18.4847 &	129.5 &	0.4 &	1.8017 &	0.4 &	0.6 *    \\ 
             &         &                  &     15.5000 &       129.8 & \ldots& 1.792  & \ldots     &   0.6  G \\ 
16130$-$2502 &	\ldots &	203760219 &	18.2524 &	152.0 &	0.9 &	0.0694 &	2.0 &	0.2      \\ 
16134$-$2726 &	\ldots &	203048597 &	18.2524 &	147.2 &	0.3 &	0.7388 &	0.3 &	0.0      \\ 
             &         &                  &     15.5000 &       145.6 & \ldots & 0.741 & \ldots     &   0.1 G      \\
16136$-$2326 &	\ldots &	204156820 &	18.2524 &	155.9 &	0.3 &	0.5471 &	0.7 &	0.0      \\ 
            &	       &	          &	18.4847 &	155.7 &	0.8 &	0.5482 &	0.8 &	0.0 :    \\ 
16137$-$1846 &	\ldots &	205225696 &	18.4847 &	63.4 &	2.8 &	0.2056 &	16.8 &	0.0 :    \\ 
16137$-$2215 &	\ldots &	204459941 &	18.2526 &	110.6 &	0.2 &	0.0586 &	0.2 &	0.4 :    \\ 
16138$-$2425 &	B 307 AB &	203917770 &	18.4818 &	217.4 &	0.4 &	1.7434 &	0.5 &	3.0 *    \\ 
             &           &                &     15.5000 &       218.0 & \ldots& 1.729  & \ldots     &   3.4 G   \\ 
16138$-$2748 &	      AB &	202947197 &	18.4817 &	271.4 &	0.9 &	0.3618 &	0.5 &	4.5      \\ 
16138$-$2748 &	      BC &	202947197 &	18.4817 &	195.3 &	0.5 &	0.0630 &	2.4 &	0.7      \\ 
16139$-$2458 &	\ldots &	203777800 &	18.3999 &	23.6 &	6.0 &	0.2043 &	2.3 &	0.1 :    \\ 
16142$-$2217 &	\ldots &	204449800 &	18.4847 &	35.6 &	1.5 &	0.8400 &	1.5 &	0.0 :    \\ 
             &         &                  &     15.5000 &       35.4 & \ldots & 0.838  & \ldots     & 0.0  G    \\
16149$-$2308 &	\ldots &	204235325 &	18.2524 &	39.6 &	0.6 &	0.6951 &	0.9 &	1.2 *    \\ 
16150$-$2920 &	\ldots &	202533810 &	18.2524 &	61.3 &	0.2 &	0.1745 &	1.4 &	0.0      \\ 
16159$-$2315 &	\ldots &	204204606 &	18.3999 &	95.7 &	1.8 &	0.2707 &	5.5 &	0.0 :    \\ 
16163$-$2344 &	\ldots &	204082531 &	18.3999 &	21.1 &	12.7 &	0.0594 &	13.8 &	0.0 :    \\ 
16175$-$2451 &	\ldots &	203809317 &	18.4846 &	308.8 &	2.5 &	0.4112 &	1.1 &	0.5 :    \\ 
16198$-$2148 &	\ldots &	204569229 &	18.2526 &	95.9 &	0.3 &	0.0510 &	0.3 &	0.4 :    \\ 
16205$-$2126 &	\ldots &	204655550 &	18.4847 &	37.5 &	3.3 &	0.0751 &	9.5 &	0.9 :    \\ 
16226$-$2235 &	\ldots &	204374147 &	18.2526 &	204.3 &	2.2 &	0.0830 &	1.3 &	0.8 q    \\ 
16240$-$2525 & Gaia    &        203649927 &     15.5000 &         8.4 & \ldots& 2.059 & \ldots      &   4.1  G   \\
16240$-$2736 &	     AB &	203001867 &	18.2524 &	145.9 &	0.0 &	0.4424 &	0.8 &	0.4 :    \\ 
            &	        &	          &	18.3999 &	146.6 &	0.1 &	0.4370 &	0.2 &	0.3 :    \\ 
16240$-$2736 &	    BC &	203001867 &	18.2524 &	293.0 &	5.2 &	0.1403 &	2.7 &	1.0 :    \\ 
            &	       &	          &	18.3999 &	291.8 &	0.2 &	0.1469 &	0.2 &	0.8 :    \\ 
16241$-$2457 &	\ldots  &	203783190 &	18.2524 &	67.7 &	0.3 &	0.2429 &	0.1 &	0.2      \\ 
16244$-$2502 &	B 308 AB &	203761347 &	18.4818 &	142.2 &	1.2 &	2.7129 &	1.2 &	4.6 *    \\ 
             &           &                &     15.5000 &       142.2 & \ldots& 2.704  & \ldots     & 4.4  G     \\   
16249$-$2214 &	\ldots &	204462113 &	18.2526 &	156.2 &	0.5 &	0.0779 &	0.5 &	0.2 :    \\ 
16249$-$2240 &	KSA 129 Aa,Ab &	204355491 &	18.4818 &	331.7 &	0.9 &	0.0432 &	1.6 &	0.5      \\ 
16251$-$2432 &	\ldots &	203888630 &	18.4846 &	103.3 &	2.4 &	0.2902 &	4.3 &	0.0 :    \\ 
16253$-$2427 &	B 309       &	203911473 &	18.4818 &	295.9 &	0.5 &	0.1381 &	0.7 &	0.0      \\ 
16259$-$2721 &	\ldots &	203071614 &	18.3999 &	159.4 &	3.1 &	0.1086 &	4.7 &	0.4 :    \\ 
16267$-$2200 &	\ldots &	204520585 &	18.2526 &	231.9 &	1.3 &	0.1071 &	6.8 &	0.9 q    \\ 
16267$-$2440 &	RAT 7       &	203856244 &	18.4846 &	78.0 &	4.8 &	0.1386 &	5.0 &	0.5 :    \\ 
16273$-$2442 &	CST 8 AB &	203850605 &	18.2524 &	77.8 &	0.3 &	0.1457 &	1.5 &	0.0      \\ 
16282$-$2451 &	\ldots &	203810698 &	18.4846 &	101.2 &	3.7 &	3.6236 &	2.7 &	1.0 :    \\ 
16288$-$2712 &	\ldots &	203115615 &	18.3999 &	149.7 &	1.7 &	0.1257 &	1.2 &	1.0 :    \\ 
16292$-$2901 &	\ldots &	202615424 &	18.3999 &	144.5 &	6.1 &	0.1275 &	2.5 &	0.0 :    \\ 
16296$-$2436 &	\ldots &	203873374 &	18.4846 &	129.2 &	7.8 &	0.1858 &	69.2 &	0.1 :    \\ 
16341$-$2659 &	\ldots &	203181641 &	18.4846 &	243.4 &	4.5 &	0.0902 &	5.7 &	1.1 q    \\ 
16352$-$2846 &	\ldots &	202679988 &	18.4846 &	63.7 &	0.3 &	0.0615 &	5.0 &	0.9      
\enddata 
\end{deluxetable*}   

\subsection{Distribution of separations}

Distribution of periods or separations of low-mass binaries is
frequently approximated by a log-normal function.  Let $ x = \log_{10}
\rho$ be the decimal logarithm of the separation $\rho$. The
log-normal distribution $f(x)$, 
\begin{equation}
f(x) = \epsilon (2 \pi \sigma)^{-1} \exp [- (x-x_0)^2/(2 \sigma^2) ], 
\label{eq:f}
\end{equation}
has  three   parameters:  multiplicity  fraction   $\epsilon$,  median
separation  $x_0$,  and the  logarithmic  dispersion $\sigma$.   These
parameters  were  determined by  the  maximum  likelihood (ML)  method
\citep[see  for example  the  Appendix of  ][]{FG67}.   We assume  the
detection probability that rises linearly in the separation range from
0\farcs04 to  0\farcs07 (inspired  by the sharpness  of the  cutoff in
Figure.~\ref{fig:sep-dm})  and  equals  one  for separations  $\rho  <
3\arcsec$ ~(binaries with larger  separations are ignored).  For stars
with  several companions,  we  use the  smallest  separation from  the
primary component (separation  between A and B in  triple systems like
A-BC and AB-C). The  derived separation distribution characterizes all
companions  regardless of  magnitude difference  $\Delta I$.  As tight
pairs    with    large    $\Delta    I$    can    be    missed    (see
Figure~\ref{fig:sep-dm}),  the distribution  can be  biased  to larger
separations.

\begin{figure}
\epsscale{1.1}
\plotone{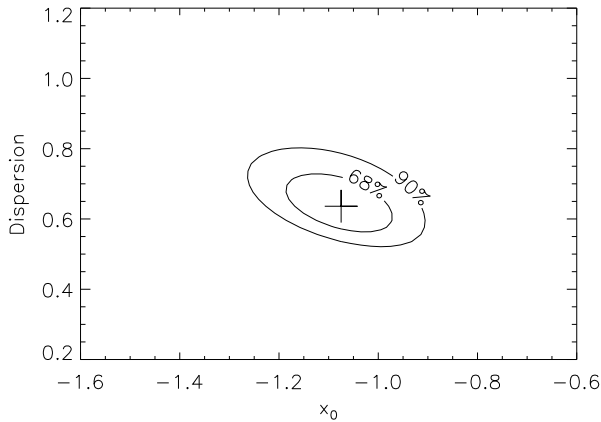}
\caption{Contours of  the likelihood function for  objects with masses
  $< 1 {\cal M}_\odot$ in the $(x_0, \sigma)$ plane that correspond to
  the significance levels of 68\% and 90\%. 
\label{fig:ML}
}
\end{figure}

The first parameter, multiplicity fraction, is naturally restricted to
the range $0 <  \epsilon < 1$. The ML fitting leads  to $\epsilon = 1$
(i.e.  all  targets are binary).   The two remaining  free parameters,
$x_0$   and   $\sigma$,  are   slightly   correlated,   as  shown   in
Figure~\ref{fig:ML}.

\begin{figure}
\epsscale{1.1}
\plotone{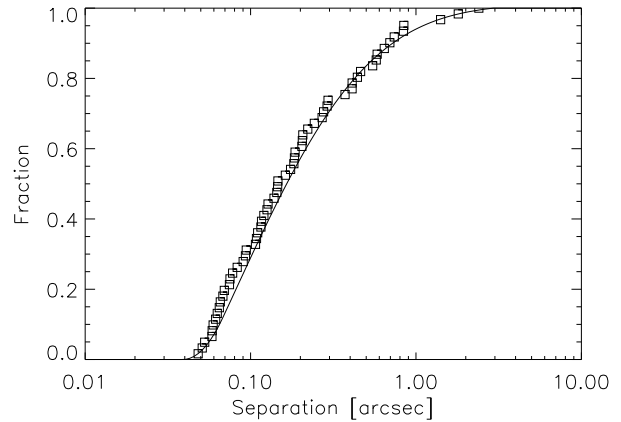}
\caption{Cumulative  distribution of  projected  separations (squares)
  and its  log-normal model  (line) for 95 binaries  less massive  than 1
  ${\cal M}_\odot$.
\label{fig:hist}
}
\end{figure}

\begin{deluxetable}{cc cc cc c  }
\tabletypesize{\scriptsize}     
\tablecaption{Separation distribution
\label{tab:ML} }  
\tablewidth{0pt}                                   
\tablehead{      
\colhead{Range} & 
\colhead{ $\langle M \rangle$} & 
\colhead{$N_{\rm tot}$} & 
\colhead{$N_{\rm bin}$} & 
\colhead{$x_0$} & 
\colhead{$\sigma$} &
\colhead{$\epsilon$} \\
\colhead{(${\cal M}_\odot$)} & 
\colhead{(${\cal M}_\odot$)} & 
  &   &  &  
\colhead{(dex)}  &
}
\startdata
$<$1   & 0.52 & 95  & 61  &   $-1.08_{-0.09}^{+0.09}$  & $0.64_{-0.06}^{+0.08}$ & $>$0.91 \\
$<$0.5 & 0.38 & 44  & 29  &   $-1.04_{-0.13}^{+0.14}$  & $0.62_{-0.08}^{+0.13}$ &  $>$0.85 \\
0.5--1 & 0.69 & 51  & 32  &  $-1.11_{-0.12}^{+0.13}$  & $0.65_{-0.08}^{+0.12}$ &  $>$0.87 \\
$>$1   & 1.65 & 34  & 9   &   \ldots                  & \ldots & \ldots  \\ 
All & 0.63  & 129 & 70 & $-1.20_{-0.11}^{+0.11}$  & $0.79_{-0.08}^{+0.11}$ & $>$0.91 
\enddata
\end{deluxetable}

The results of parameter-fitting for sub-samples in selected ranges of
mass are given in Table~\ref{tab:ML}.  Its first column gives the mass
range, the second column is the median mass in this range. Then follow
the  total number  of  observed  targets and  the  number of  resolved
binaries  (ignoring  pairs with  $\rho  >  3$\arcsec).  The  remaining
columns  give the  best-fit parameters  $x_0$ and  $\sigma$  and their
formal  errors   corresponding  to  68\%   confidence  intervals  (``1
$\sigma$'').  Figure~\ref{fig:hist}  shows the cumulative distribution
of projected separations  for the $< 1 {\cal  M}_\odot$ sub-sample and
its ML model. At the  smallest separations, the distribution is shaped
by  the  detection  limit  (note  the  initial  quadratic  rise).  The
log-normal model fits the data quite well.

We see that the log-normal distributions approximating the full sample
and its sub-groups are  mutually consistent, showing robustness of the
result.  The raw fraction of  resolved binaries is always close to 2/3
and the  derived multiplicity fraction  is always $\epsilon =  1$ (the
68\%   confidence   lower   limits   of  $\epsilon$   are   given   in
Table~\ref{tab:ML}).   Only  massive  targets  with  $  M  >  1  {\cal
  M}_\odot$ stand apart by the low fraction of resolved binaries, 0.26.
The ML  fitting for  this group does  not yield  meaningful parameters
owing to the small number of  binaries.  We do not discuss these stars
further and focus only on 95 targets with $ M < 1 {\cal M}_\odot$.

Assuming the  distance of  140\,pc, the median  separation is  11.6 au
($x_0  = 1.07$  au).  The  logarithmic  dispersion is  0.6 dex.   Both
parameters do  not vary significantly  between the low-  and high-mass
parts of  the sample.  We  caution that the formal  errors of the
log-normal parameters  are meaningful only when  the true distribution
is indeed log-normal.

Let us compare the  log-normal separation distribution found here with
prior     knowledge     on      the     statistics     of     low-mass
binaries. \citet{Janson2012} surveyed early M-type dwarfs in the solar
neighborhood and found that  their semimajor axes match the log-normal
distribution with  $x_0 = 1.2$ au  and $\sigma = 0.8$  dex.  This team
also observed  dwarfs of  spectral types later  than M5V in  the field
\citep{Janson2014} and  derived the more  compact distribution shifted
to smaller separations:  $x_0 = 0.78$ au and $\sigma  = 0.47$ dex.  On
the other hand, solar-mass binaries  in the field have a larger median
separation $x_0 = 1.7$ au and a larger dispersion $\sigma = 1.32$ dex
\citep{FG67}. 

The  multi-periodic  stars  in  USco  observed  here  have  separation
distribution similar to that of  field M dwarfs. However, the trend of
decreasing  median  separation and  dispersion  with decreasing  mass,
observed in the  field, is not found in  this sample; the distribution
of binary separations in USco appears to be independent of mass in the
range from 0.3 to 1 ${\cal M}_\odot$.

\subsection{New triple stars}

The lower panels of Figure~\ref{fig:ACF} illustrate new triple systems
discovered in this program.  The  first one, EPIC 20429883, is listed
in WDS  as J16093$-$2222 or WSI~130,  but only as  a 1\arcsec ~binary.
Its discovery in 2010 was announced recently by \citet{Mason2018}.  We
detected the inner close pair Aa,Ab composed of two equal stars, while
the  previously known  distant companion  B is  substantially fainter,
hence less massive. Interestingly,  RSC18 found 4 distinct photometric
periods for  this star.  It  was observed at  SOAR in 2015.3,  but the
triple nature was overlooked at the time. We include the 2015 measures
in Table~\ref{tab:double}.  The inner pair Aa,Ab turned by 10\degr ~in
3 years. 

Another triple star EPIC~202947197  (J16138$-$2748) was discovered in
June.  A 0\farcs06 pair of faint stars B and C is located at 0\farcs36
from the main star A, which  has a relatively large mass of 1.5 ${\cal
  M}_\odot$.   RSC18  suspected  a  disk and  detected  4  photometric
periods.

The  third  triple  system  EPIC~203111867 (J16240$-$2736)  is  a  new
discovery made in  April and confirmed by re-observation  in May. Here,
the  close   0\farcs14  pair  B,C   orbits  the  central  star   A  at
0\farcs44. Note that the ratio  of separations is small, only 3.1. So,
this  young  triple  system  could  be  dynamically  unstable,  unless
comparable separations result from projection  on the sky and the real
ratio is larger.  RSC18 detected only two periods.

Our sample  contains at least  two unresolved compact  triple systems.
One  of  those, EPIC  204506077  (HD~144548),  is  a triply  eclipsing
system  with periods  of 33.9  and 1.63  day  \citep{Alonso2015}. The
directly  measured mass of  the primary  star, 1.44  ${\cal M_\odot}$,
compares  well with  1.67  ${\cal M_\odot}$  estimated  here from  the
isochrone. Another  triple is EPIC  203476597, where \citet{David2016}
found  a  1.4 day  eclipsing  binary  accompanied  by a  more  massive
star. For both triples, RSC18 detected only two photometric periods.

There  are triple  systems with  outer companions  outside  the survey
radius of 3\arcsec  ~listed in the WDS (EPIC  204844509 and 203937317)
and three new triple systems  with wide companions found by {\it Gaia}
(EPIC 204179058, 204637622, and  204649937). For several other objects
in  Table~\ref{tab:sample},   RSC18  found  three   or  four  distinct
photometric periods that could be  produced by additional stars in the
systems.  


\subsection{Orbits and masses}

\begin{figure}
\plotone{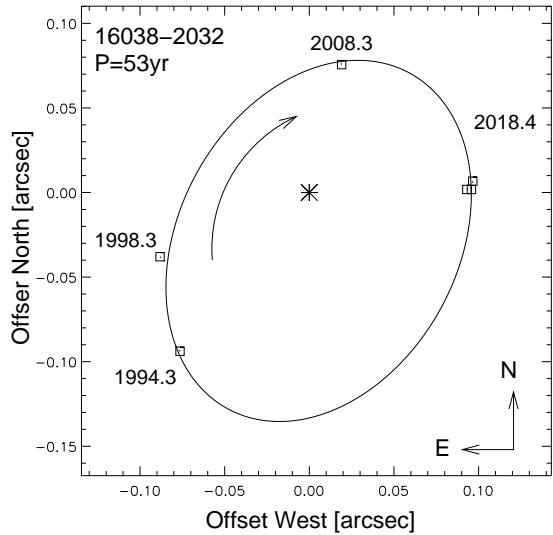}
\caption{Preliminary orbit of EPIC 204862109 (WDS J16038-2032, KOH 70)
  with  a  period   of  53  years.  Squares  with   dates  denote  
  measurements  of the  secondary companion  relative to  the primary,
  placed at the coordinate origin.
\label{fig:orb}
}
\end{figure}

With a separation of 10\,au and  a mass sum of 1 ${\cal M}_\odot$, the
orbital period of  a typical binary is around 30  years.  We note that
the positions  of some  previously known close  binaries re-discovered
here  differ  from the  published  measures,  owing  to their  orbital
motion.  One of the brightest pairs in our sample, EPIC 204862109 (WDS
J16038-2032, KOH  70), has  3 prior measures  dating back to  1994 and
shows considerable  orbital motion. Combining these data  with our own
measures,    we    computed     a    tentative    orbit    shown    in
Figure~\ref{fig:orb}. Its elements are: $P = 52.6$ year, $T_0 = 2006.3$,
$e=0.3$, $a=  0\farcs116$, $\Omega  = 147\fdg6$, $\omega  = 127\fdg5$,
and $i  = 136\fdg4$. Using the  DR2 parallax of  6.59$\pm$0.22 mas, we
compute the  mass sum of 1.7 ${\cal  M}_\odot$.  Considering potential
bias  of  the  DR2 parallax  (its  large  error  is likely  caused  by
binarity)  and the intrinsic  uncertainty of  masses derived  from the
isochrones,  the estimated mass  sum of  1.5 ${\cal  M}_\odot$ matches
quite well  the mass derived from the orbit.  

Further  monitoring  of  the  orbital  motion and  a  better  parallax
measurement from  forthcoming {\it Gaia} data releases  will result in
accurate  masses that will  help to  calibrate evolutionary  tracks of
young  stars.  We  expect that  future {\it  Gaia} data  releases will
detect acceleration of these stars (including closer binaries that are
not resolved here).   New close pairs discovered here  are amenable to
orbit determination if  their motion is monitored in  the future. 


\section{Discussion}
\label{sec:disc}

The  large   number  of  newly  resolved   binaries  appears  somewhat
surprising, considering that  USco has received considerable attention
from  this  perspective. The  pioneering  work by  \citet{Koehler2000}
surveyed 118 stars in USco with an angular resolution of 0\farcs13 and
revealed  many  binaries (hence  the  discoverer  codes  KOH in  WDS).
\citet{Kraus2007} used  seeing-limited 2MASS imagery  \citep[see also
  the  follow-up  work  by][]{Aller2013}.  \citet{Kraus2008}  explored
USco  stars brighter  than $R=14$  mag and  north of  $-25^\circ$ with
high-contrast  aperture  masking,   excluding  known  binaries,  while
\citet{Kraus2012} used laser adaptive  optics to probe binarity of the
78   lowest-mass  members  of   USco.     \citet{Janson2013}  and
  \citet{Laf2014}  observed members  of USco  not covered  by previous
  high-resolution imaging.

These  high-resolution studies  involved  modest samples  and did  not
cover  the complete  USco  association.  As  most  low-mass stars  are
single, a  sample of $\sim$100 targets yields  only $\sim$20 binaries.
In contrast, we start here with multi-periodic stars selected from the
large parent sample.   Our targets are from the  outset expected to be
binaries  with  components  of  comparable flux.   This  pre-selection
ensures the large fraction of resolved binaries we found.  With only a
few hours of telescope time, we could substantially enlarge the number
of known binaries  among low-mass members of USco.   Many  new tight
pairs  have short  periods and  are an  excellent material  for future
calibration of PMS stellar evolutionary models.

It will  be interesting to test binarity  of unresolved multi-periodic
targets   using  either   a   higher  spatial   resolution  and/or   a
high-resolution   spectroscopy.    Both   approaches   involve   large
telescopes, hence pre-selection of  unresolved targets from this study
would be a valuable starting point allowing to save telescope time. We
expect that {\it  Gaia} will resolve many of  these stars, will detect
their  accelerations, and  will  determine astrometric  orbits of  the
closest and  fastest pairs. Lunar occultations  is another potentially
interesting  method  of  probing  binarity  in USco,  although  it  is
applicable only to relatively bright stars.

\acknowledgements

We are grateful to J.~Stauffer who suggested to look at multi-periodic
stars using  speckle interferometry and made valuable  comments on the
early  versions of  this paper.   This  work used  the SIMBAD  service
operated  by  Centre des  Donn\'ees  Stellaires (Strasbourg,  France),
bibliographic references from  the Astrophysics Data System maintained
by  SAO/NASA, and  the Washington  Double Star  Catalog  maintained at
USNO.

{\it Facilities: SOAR}










\end{document}